\documentclass{article}       
\usepackage{epsf}            
\def\roots{\sqrt{s}}

\def\z0{\rm Z^0}
\def\as{\alpha_{\rm s}}

\newcommand{\oaa}{{\cal O}(\as^2)}
\newcommand{\oaaa}{{\cal O}(\as^3)}
\newcommand{\epem}{{\rm e^+\rm e^-}}
\newcommand{\yc}{y_{\rm cut}}
\newcommand{\amz}{\as(M_{\rm Z^0})}

\def\mz{M_{\rm Z^0}}

\def\rz{R_{\rm Z}}
\def\d2{D_2}
\def\oq{\char'134}

\def\ecm{E_{cm}}

\def\m2{\mu^2}
\def\q{\rm q}

\def\q2{Q^2}
\def\asq{\as (\q2 )}

\def\mtau{M_{\tau}}
\def\r3{R_3}
\def\gcc{g_{c\bar{c}}}
\def\gbb{g_{b\bar{b}}}
\begin{document}
\title{QCD Studies at LEP}
\author{S. Bethke \\
Max-Planck-Institute of Physics, 
80805 Munich, Germany }
\maketitle
\begin{abstract}
Studies of hadronic final states of $\epem$ annihilations at LEP are reviewed.
The topics included cover hadronic event shapes, measurements of $\as$, 
determinations of QCD colour factors and tests of the non-Abelian 
gauge structure of QCD, differences between quark and gluon jets, 
QCD with heavy quarks
and selected results of two-photon scattering processes.
\end{abstract}
\section{Introduction}
The  
LEP experiments ALEPH \cite{a}, DELPHI \cite{d}, L3 \cite{l} 
and OPAL \cite{o} have contributed more than 240 publications on hadronic 
physics and tests of Quantum Chromodynamics (QCD),
the theory of the Strong 
Interaction between quarks and gluons (see e.g. \cite{qcd}).
On the occasion of the $50^{th}$ anniversary of the CERN laboratory 
in October 2004, four years after the close-down of the LEP collider,
this article gives an overview of some of the main QCD results 
at LEP.

The emphasis of this review is concentrated on 
studies which, based on perturbation theory, test key features of QCD.
For earlier reviews of hadronic physics at LEP, the reader is referred to 
\cite{betpil,hebbecker,schmelling,stenzel,sb-lepfest}.

\section{Hadronic Events at LEP}
According to the current understanding 
of high energy particle collisions and reactions
in the framework of the Standard Model, see e.g.  \cite{sm,lepewwg03,sm-lep},
hadronic final states in $\epem$ annihilations are produced 
through an intermediate virtual photon or a $\z0$ boson,
which decays into a quark-antiquark pair.
The development of a quark-gluon cascade from the 
initial quark-antiquark system is calculated in fixed order QCD 
perturbation theory, so far in full next-to-leading order 
(NLO, equivalent to $\oaa$) \cite{ert,kn,catsey}, or in the (next-to-)leading 
logarithmic approximation ((N)LLA) \cite{nlla}. 
The nonperturbative process of 
hadronisation into visible particles is described by phenomenological string- 
\cite{string} or cluster- \cite{cluster} 
fragmentation models or, alternatively, 
by applying analytical power corrections \cite{powcor}.

At c.m. energies above the threshold of $W$- or $\z0$-boson 
pair production, hadronic final states are also generated through the 
decays of these bosons to 4 fermions, if at least one of these pairs is a 
quark-antiquark system.
The physics of 4-fermion final states is not included in this review but is
discussed elsewhere 
\cite{lepewwg03}.

\begin{figure}[htb]
\begin{center}
\epsfxsize16.0cm
\epsffile{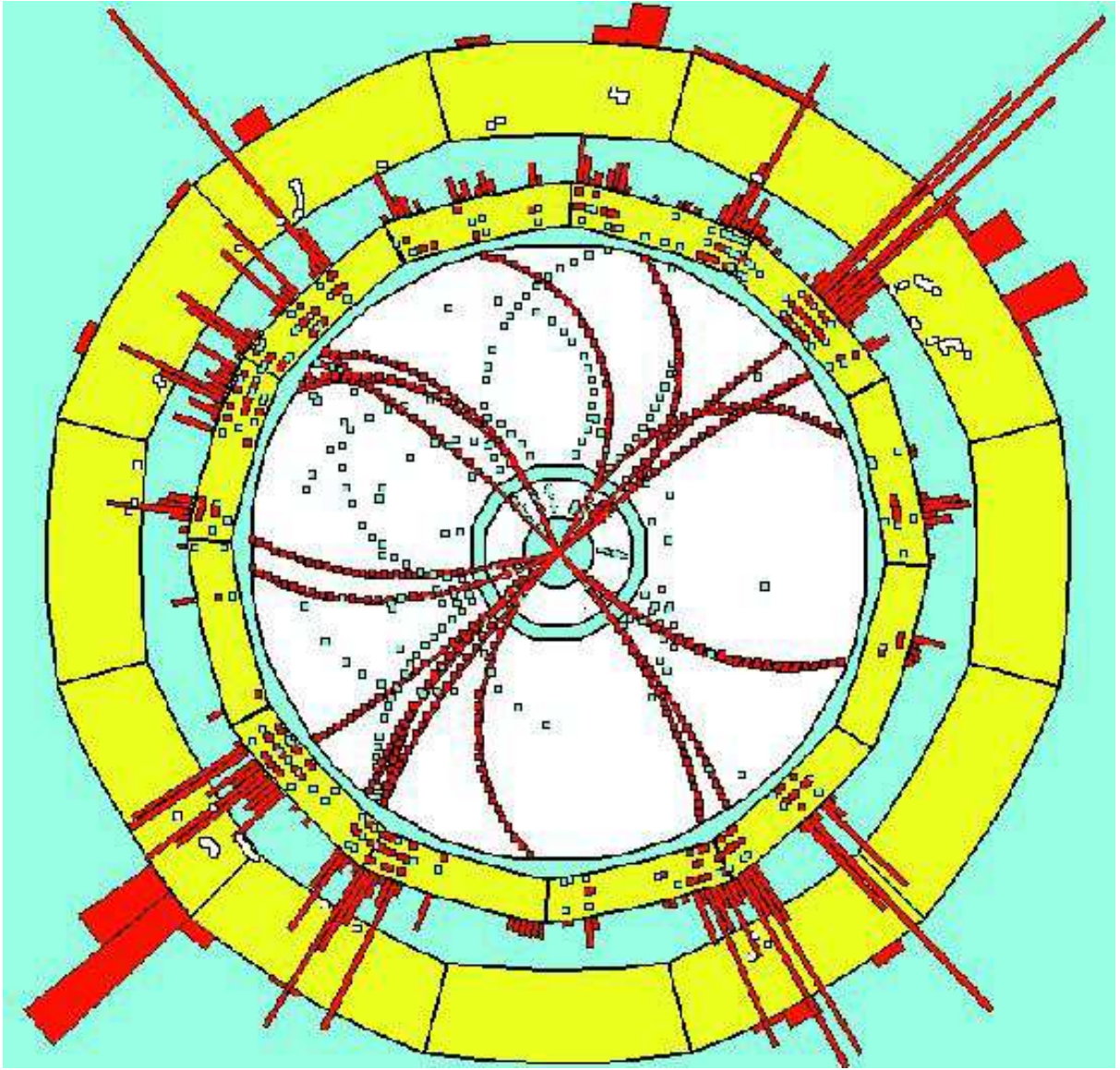}
\end{center}
\caption{\label{A-4jet-event}
Hadronic event of the type $\epem \rightarrow {\rm 4\ jets}$
recorded with the ALEPH detector at LEP-I.}
\end{figure}

During its time of physics operation, from August 1989 to November 2000, 
the LEP collider delivered an integrated luminosity of about 1~$fb^{-1}$ 
to each of the four experiments.
Of this, about 200~$pb^{-1}$ were collected during the \oq LEP-I" phase
of operation, from 1989 to 1995, 
at or around the $\z0$ mass resonance, i.e. at 
$E_{cm} \sim \mz = 91.1875 \pm 0.0021\ {\rm GeV}$ \cite{lepew-0404}.
This, together with the large resonant $\epem$ annihilation cross section 
at the $\z0$ mass, resulted in data samples of about 
4 million hadronic events for each experiment. 
A typical example of an event $\epem \rightarrow {\rm 4\ jets}$ is
shown in Figure~\ref{A-4jet-event}.

The \oq LEP-II" phase, from 1996 to 2000 at c.m. energies at and above the
pair-production of $W$ bosons, up to a maximum of 209~GeV, resulted in
integrated luminosities of about 750~$pb^{-1}$. 
The approximate total numbers of hadronic events, obtained by each LEP 
experiment, are summarised in Table~1.

Due to the large event statistics, the clean and precise 
environment of $\epem$ annihilations, the high c.m. energies, the improved 
detector technology and advanced theoretical calcluations,
significant achievements were achieved at LEP, compared to the 
time before, see e.g. \cite{sb-budapest,altarelli-as,sb-lepfest}.

\section{Hadronic Event Shapes and Jet Production}
At the time of LEP operation, measurements of jet 
production rates and of hadronic event shape parameters developed into 
precision tools to determine $\as$, to probe details of perturbative QCD 
predictions, to study hadronisation properties and to optimise 
and test hadronisation models.
The development was largely influenced by 
the introduction of new jet algorithms \cite{bkss}, most notably
the Durham (D-) scheme algorithm \cite{d-jets}, of new event shape
measures \cite{broadening} and of improved theoretical predictions
\cite{catsey,nlla,nlo-4jet}.
Overviews of jet and event shape observables can be found e.g. in
\cite{kn,bkss,nijmegen}.

\begin{figure}[htb]
\begin{center}
\epsfxsize12.0cm
\epsffile{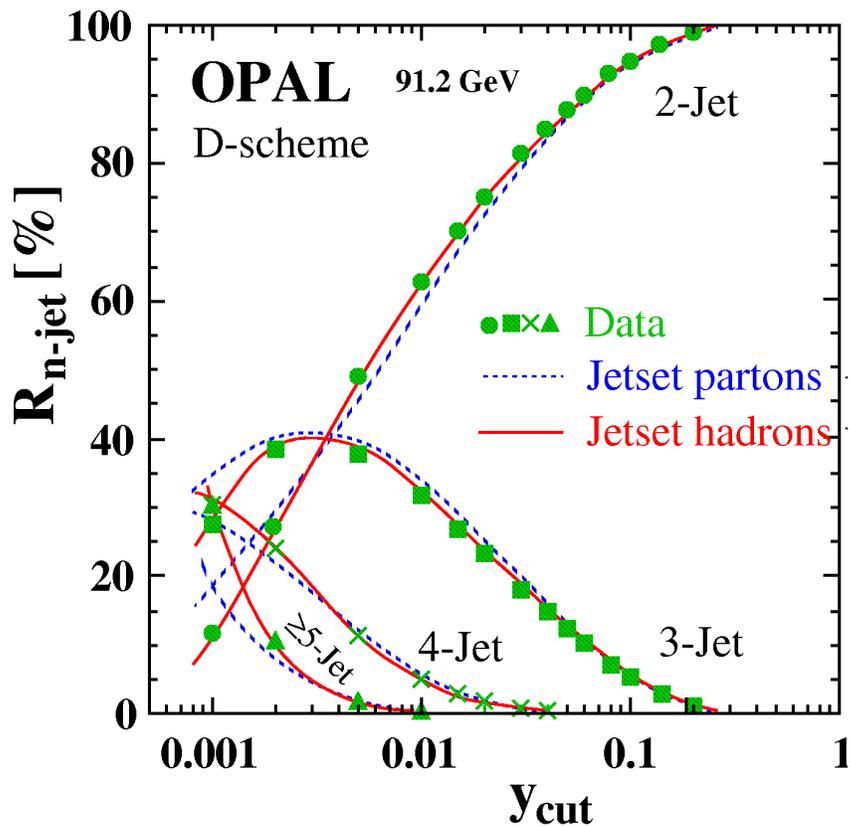}
\end{center}
\caption{\label{O-djets}
Relative production rates of n-jet events (n = 2 to 5) for different values of
the jet resolution parameter $\yc$, measured at the
$\z0$ resonance at LEP \cite{O-global}.
The data are compared to predictions of the JETSET QCD shower and hadronisation
model ({\it hadrons}). 
The predictions for {\it partons}, before hadronisation, are
also given in order to illustrate the size of the hadronisation effect.}
\end{figure}

\begin{figure}[htb]
\begin{center}
\epsfxsize12.0cm
\epsffile{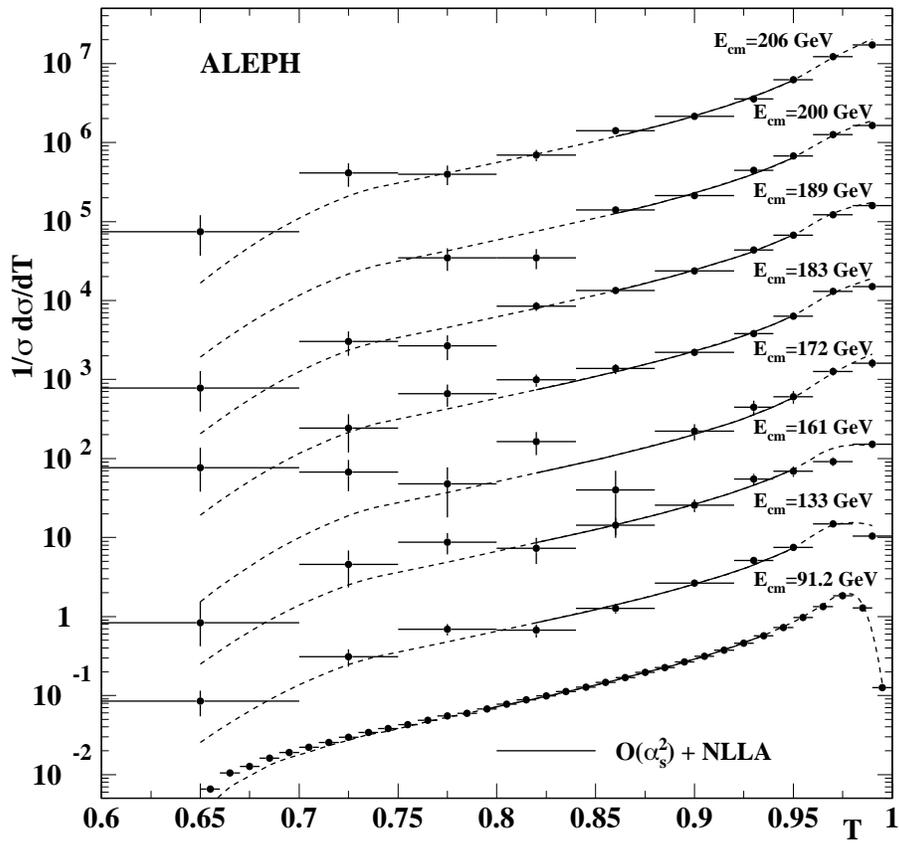}
\end{center}
\caption{\label{A-t}
Measured distributions of Thrust \cite{A-new-shapes}, after corrections for
backgrounds and detector effects, together with fitted QCD predictions.}
\end{figure}

The precision of data description by QCD model calculations is exemplified in 
Figure~\ref{O-djets}, where the measured 
relative production rates of multijet 
events are compared to the predictions of QCD shower models, at $\ecm = 91.2\ {\rm GeV}$
\cite{O-global},
and in Figure~\ref{A-t}, 
where the distributions of the shape observable Thrust (T)
\cite{thrust}, measured at different c.m. energies, are compared
with analytical predictions of QCD \cite{A-new-shapes}.
QCD shower models as well as
QCD analytical predictions, with their parameters optimised
to provide an overall good description of the data, are able to reproduce 
even subtle dynamic features of the data, over the entire LEP energy range.
Hadronisation effects are, for many observables, small and well
under control. 

\section{Determinations of $\as$}
The coupling parameter of the Strong Interactions, $\as$, is - similar to 
the fine structure constant $\alpha_{em}$, the Weinberg angle $sin^2 \theta_w$ 
and the mass of the electron $m_e$ - one of the basic constants of
nature, whose values, however, are not given by theoretical
predictions but must be measured by experiment.
Precise measurements of $\as$ and the experimental verification of
the energy dependence of $\as$, specifically as predicted by QCD (see
e.g. \cite{qcd,concise,as2002}), therefore were (and still are) one of the
key research issues at LEP.

\subsection{$\as$ from Electroweak Precision Measurements}
Determinations of $\as$ from electroweak precision measurements
crucially depend on the strict validity of the predictions
of the Standard Model.
QCD corrections affect almost all electroweak precision observables and
measurements at LEP.
In particular, the hadronic partial decay
width of the $\z0$, $\Gamma_{had}$, 
obtains QCD corrections of the form $\left( 1 + \Sigma_n ( C_n \as^n )\right)$, 
$n = 1, 2, 3, ...$.
These corrections are known up to next-next-to-leading order (NNLO), 
i.e. to $\oaaa$ or
$n = 3$ \cite{rznnlo}; see also \cite{rz-report} and references quoted therein. 

In the most recent combination of the LEP-I and LEP-II 
measurements of all four experiments,
the LEP Electroweak Working Group (LEP EWWG) \cite{lepewwg03,lepew-0404},
see also \cite{sm-lep},  obtained 
$$\amz = 0.1226 \pm 0.0038 \ (exp.) \ ^{+0.0033}_{-0.0000} \ (M_H) \
^{+0.0028}_{-0.0005} \ (QCD) $$
from $\rz = \Gamma_{had} / \Gamma_{\ell} = 20.767 \pm 0.025$, 
whereby the second error accounts for variations
of the unknown Higgs boson mass between 100 and 900 GeV/c$^2$. 
The third error 
comes from a parametrisation of the unknown higher order QCD corrections,
i.e. from variations of the QCD renormalisation scale 
and renormalistion scheme \cite{concise}.

In the same analysis \cite{lepew-0404}, 
the fitted leptonic pole cross section,
$\sigma_{\ell}^0  = (2.0003 \pm 0.0027)$~pb, resulted in
$$\amz = 0.1183 \pm 0.0030 \ (exp.) \ ^{+0.0026}_{-0.0000} \ (M_H) \ .$$
Since $\sigma_{\ell}^0  = {{12 \pi}\over{M_Z^2}} {{\Gamma_{\ell}^2}
\over {\Gamma_Z^2}}$ and $\Gamma_Z \sim \Gamma_{had}$,
$\sigma_{\ell}$ has a steeper dependence on $\as$ than has $\Gamma_{had}$: in next-to-leading order, 
the QCD coefficient $C_1$ for $\Gamma_{had}$ turns to
$2 C_1$ for $\sigma_{\ell}$, $C_2$ turns to $(2 C_2 + C_1^2)$ etc.
The experimental error of $\as$ from $\sigma_{\ell}$ is thus smaller than
that from $\Gamma_{had}$. 
However, with increased QCD-coefficients $C_i$, the renormalisation
scale uncertainty also increases, c.f. 
equation~13 of \cite{concise},
such that the QCD uncertainty on $\as$ from $\sigma_{\ell}$
is expected to roughly double w.r.t. $\as$ from $\rz$.

A global fit of all LEP data to determine
$\as$ together with the masses of the $\z0$ boson, 
of the top-quark and of the Higgs boson, gives \cite{lepew-0404}
$$\amz = 0.1200  ^{+0.0031}_{-0.0029} \ (exp.)\ .$$
The latter result is the most precise available from 
combined electroweak fits of the LEP data.
There is no additional uncertainty due to the unknown Higgs mass.
The QCD uncertainties for this particular result of $\as$, however, 
were never determined, and prove to be difficult to be guessed due
to the unknown size of the effective QCD coefficients
that enter the overall fit.
Similar as argued in the case of $\sigma_{\ell}^0$,
the QCD uncertainty on $\Gamma_{had}$ cannot simply be applied to 
other observables.

\subsection{$\as$ from $\tau$ lepton decays}
The most significant determination of $\as$ at small energy scales
is obtained from the normalised hadronic branching fraction of $\tau$ 
leptons,
$R_{\tau} = \frac{\Gamma (\tau \rightarrow {\rm hadrons}\ \nu_{\tau})} {\Gamma (\tau
\rightarrow {\rm e} \nu_e \nu_{\tau})}$,
which is predicted as~\cite{braaten}
$ R_\tau = 3.058
( 1.001 + \delta_{\rm pert} + \delta_{\rm nonpert})$.
Here, $\delta_{\rm pert}$ and $\delta_{\rm nonpert}$ are perturbative and
nonperturbative QCD corrections; $\delta_{\rm pert}$ was calculated to
complete $\oaaa$ \cite{braaten,lediberder92}
and is similar to the perturbative prediction for $\rz$.

L3 \cite{L-tau} determined $\as$ from measured branching fractions 
of tau leptons into electrons and muons.  
ALEPH \cite{A-tau} and OPAL \cite{O-tau} also presented
measurements of the vector and the axial-vector contributions to the differential
hadronic mass distributions of $\tau$ decays, 
which allow 
simultaneous determination of
$\as$ and of the nonperturbative corrections.
The latter were parametrised in terms of the operator
product expansion (OPE) \cite{OPE}.
They were found to be small and to largely cancel in the total sum of
$R_{\tau}$, as predicted by theory~\cite{braaten}.
$\as (M_{\tau})$ is 
obtained for different variants of the NNLO QCD predictions 
\cite{braaten,tau-4,rcpt}.
The combined result of $\as$ from $R_{\tau}$ (c.f. \cite{concise}) is 
$$
\as (M_{\tau}) = 0.322 \pm 0.005 {\rm (exp.)} \pm 0.030 {\rm (theo.)}\ . 
$$
When extrapolated to the energy scale $\mz$, this results
in $\amz = 0.1180 \pm 0.0005 {\rm (exp.)} \pm 0.0030 {\rm (theo.)}$.

\subsection{$\as$ from event shape observables}
Determinations of $\as$ from hadronic event shape observables, from jet
production rates and related observables are based on pure 
QCD predictions. 
They do not depend on the assumption of strict validity 
of the Standard Model, however they require  
assumptions on or parametrisations of nonperturbative
hadronisation effects.

QCD predictions for distributions and for mean values 
of hadronic event shapes, of jet
production rates and of energy correlations are available in complete NLO
\cite{ert,kn,catsey}.
In addition, for many observables, resummation of the leading and
next-to-leading logarithms (NLLA) is available \cite{nlla} which can be
matched to the NLO expressions (resummed NLO).

\begin{figure}[htb]
\begin{center}
\epsfxsize12.0cm
\epsffile{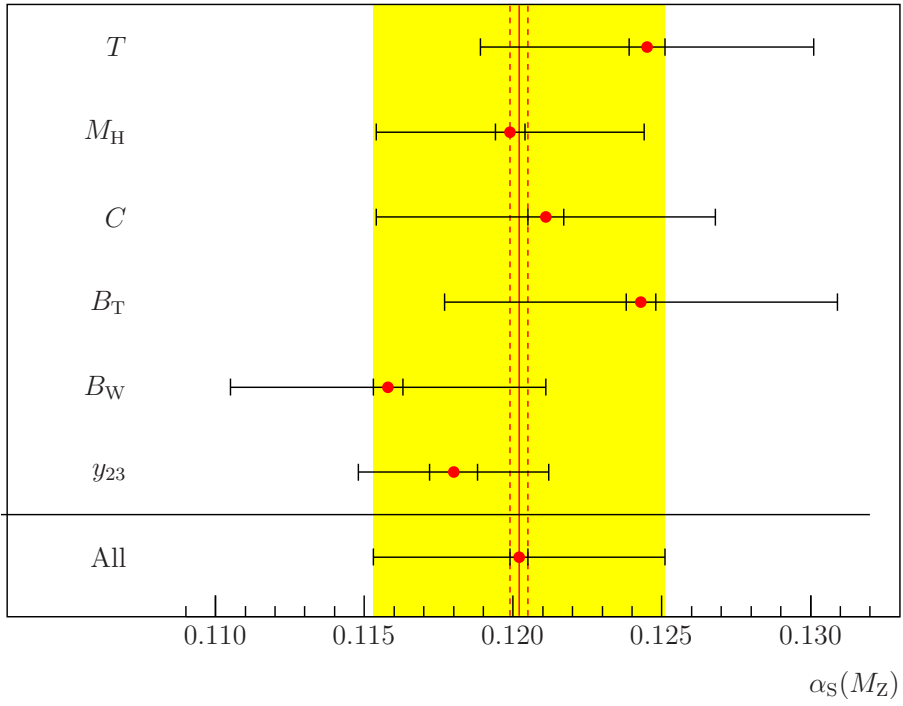}
\end{center}
\caption{\label{fitsbyvariable}
The combined $\amz$ obtained from diffent observables at LEP \cite{lepqcdwg2}.
The shaded band represents the overall combined fit for all
observables.
The inner error bars and the dashed band represent the statistical 
uncertainties.}
\end{figure}

All LEP experiments have contributed 
studies which are based on hadronic event shape observables, at all
major LEP energies, see \cite{concise,as2002} and references quoted
therein.
The LEP QCD working group has recently provided an overall
combination of all respective LEP results which is based on applying
common experimental
procedures, consistent theoretical predictions and definitions of 
the theoretical uncertainties \cite{lepqcdwg1,lepqcdwg2}.
For each observable and each energy a combined value of $\as$ is obtained.
The results for different observables are displayed 
in Figure~\ref{fitsbyvariable},
demonstrating the neccessity for a careful treatment 
and application of theoretical 
uncertainties to obtain a consistent and compatible situation.
The results of $\as$ combined for all major LEP
c.m. energies are given in Table~2.
The overall combination of all these results finally gives
$$\amz = 0.1202 \pm 0.0003\ {\rm (stat.)} \pm 0.0049\ {\rm (syst.)}\ .$$

\begin{figure}[htb]
\begin{center}
\epsfxsize12.0cm
\epsffile{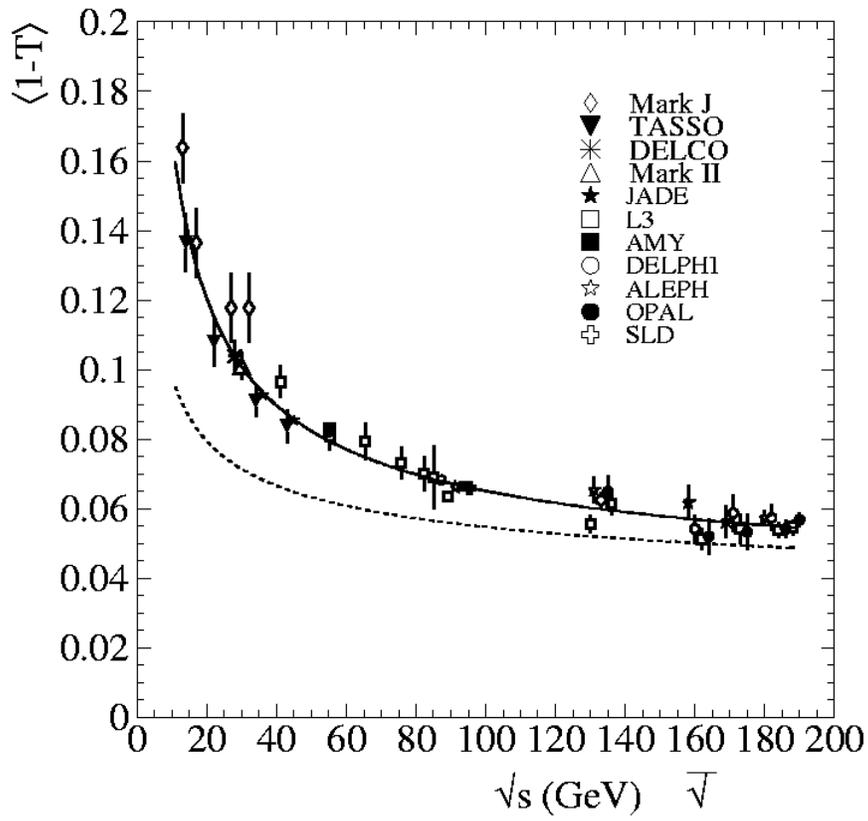}
\end{center}
\caption{\label{mean-T}
Mean values of Thrust $T$ as a function of the c.m. energy
$\roots$. 
The full line shows the QCD fit including power corrections,
the perturbative part of which is indicated by the
dashed line \cite{fernandez}.}
\end{figure}

\begin{figure}[htb]
\begin{center}
\epsfxsize12.0cm
\epsffile{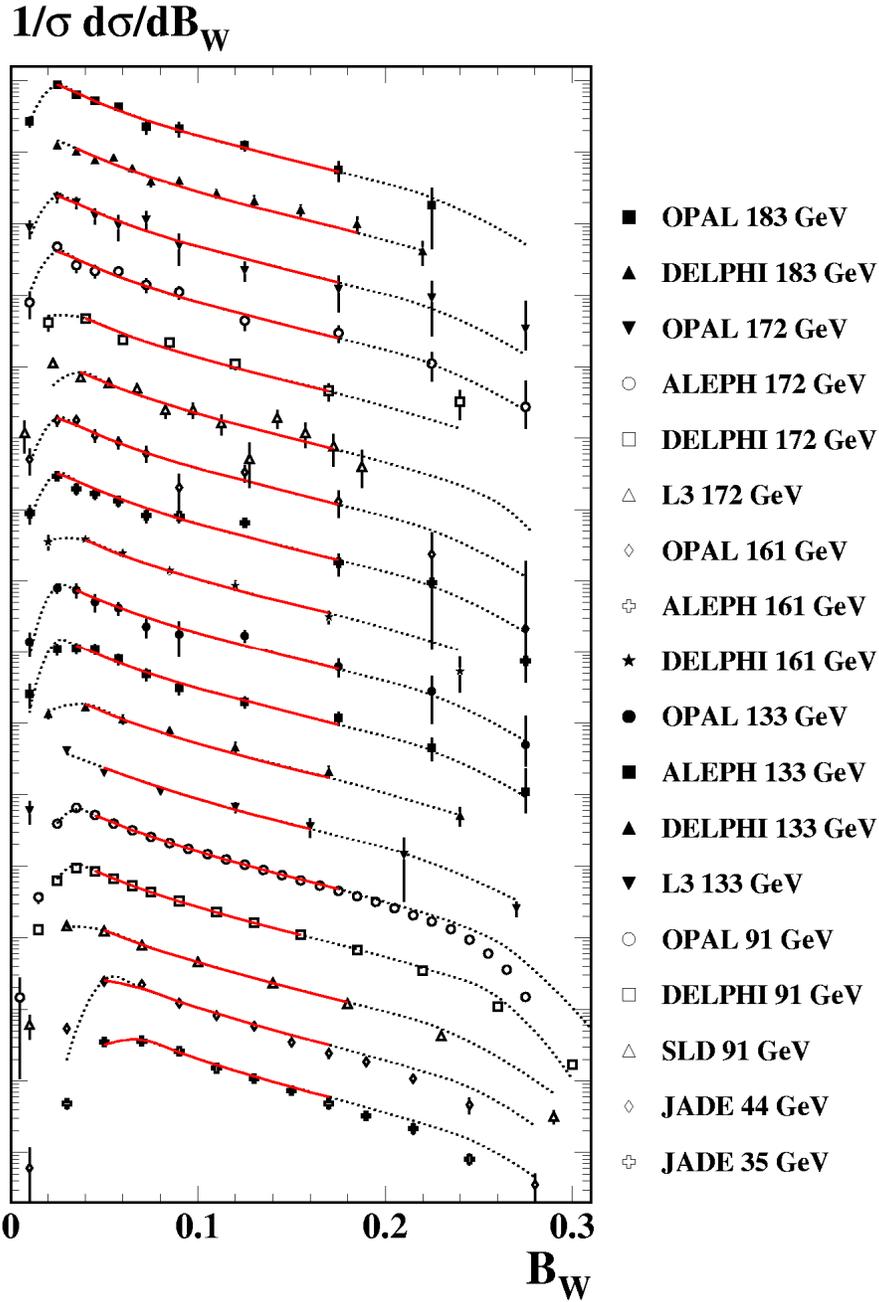}
\end{center}
\caption{\label{pedro_jetbroad}
Differential distributions of the wide jet broadening 
$B_w$ at different c.m. energies. 
The dotted lines show a common QCD fit including power corrections.
Full lines indicate the fit ranges used to adjust
$\as$ and $\alpha_0$ \cite{fernandez}.}
\end{figure}

Analytical approaches to approximate nonperturbative 
hadronisation effects lead to \oq power
corrections" which are proportional to powers of $1/Q$ \cite{powcor}.
These include, in addition to $\as$, 
only one further parameter $\alpha_0$ which stands for the unknown behaviour
of $\as$ below an infrared matching scale $\mu_I$.
Both the energy dependence of mean values as well as differential 
distributions of hadronic event shapes, without applying 
corrections for hadronisation effects, are well described 
by analytic predictions based on NLO QCD plus power corrections, see
Figures~\ref{mean-T} and~\ref{pedro_jetbroad} \cite{fernandez}.

\begin{figure}[htb]
\begin{center}
\epsfxsize12.0cm
\epsffile{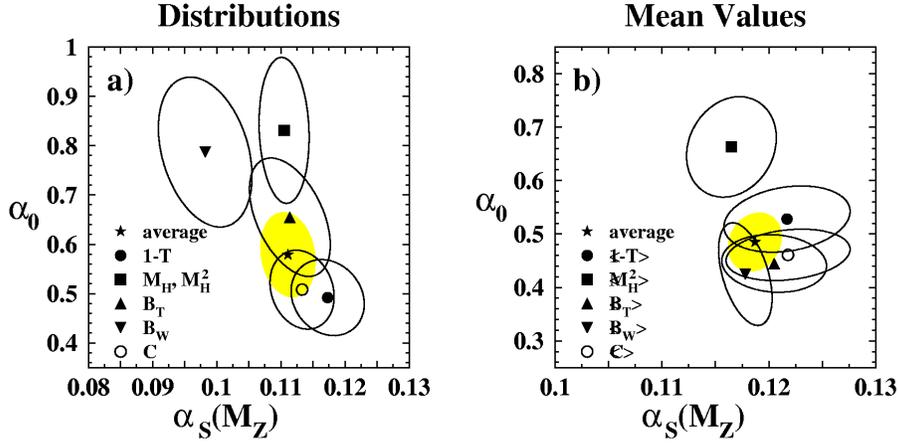}
\end{center}
\caption{\label{powcor-results}
Combined results of $\as$ and $\alpha_0$ 
from fits to the mean values and to the differential distributions
of event shape observables, measured at LEP and at lower c.m. energies
\cite{fernandez}.}
\end{figure}

A summary of fit results of $\as$ and of $\alpha_0$ \cite{fernandez} is given in 
Figure~\ref{powcor-results}.
The combined results on $\as$ from power correction fits are
$$\amz = 0.1187 \pm 0.0014\ {\rm (fit)} \pm 0.0001\ 
{\rm (sys.)} ^{+0.0025}_{-0.0015}\ 
{\rm (theo.)}$$
from mean values, and
$$\amz = 0.1111 \pm 0.0004\ {\rm (fit)} \pm 0.0020\ 
{\rm (sys.)} ^{+0.0044}_{-0.0031}\ 
{\rm (theo.)}$$
from distributions \cite{fernandez}.
The large systematic difference between these two results indicates the
presence of large but yet unknown corrections which are a matter of further 
studies.

\subsection{Other $\as$ results from LEP}
There are further studies of $\as$ from LEP, which however have not
yet reached the same experimental maturity, in terms of multiple
verification by all experiments, of the range of different systematic
checks and of verifications of the limited overall uncertainties.
These are e.g. determinations of $\as$ from studies of scaling violations 
of fragmentation functions from ALEPH~\cite{A-scaling} and 
DELPHI~\cite{D-scaling}, which can be combined
to \cite{concise}
$$ \amz = 0.125 \pm 0.007\ {\rm (exp.)} \pm 0.009\ {\rm (theo.)}\ .$$

Another notable result is the determination of $\as$ from 4-jet event production rates~\cite{A-4jet}, which is based on a NLO, i.e. $\oaaa$ QCD 
prediction~\cite{nlo-4jet}.
The 4-jet event production rate is proportional to
$\as^2$ in LO QCD, compared to $\as$ for 3-jet like shape observables, 
and thus is more sensitive to $\as$.
ALEPH obtains, with a rather rigorous definition of errors,
$$\amz = 0.1170 \pm 0.0001\ {\rm (stat.)} \pm 0.0013\ {\rm (sys.)}\ .$$

Further results on $\as$ are obtained in fits of the QCD group constants
and studies of the nonabelian nature of QCD, which are reviewed in 
section~6.

\subsection{LEP summary of $\as$}
\begin{figure}[htb]
\begin{center}
\epsfxsize12.0cm
\epsffile{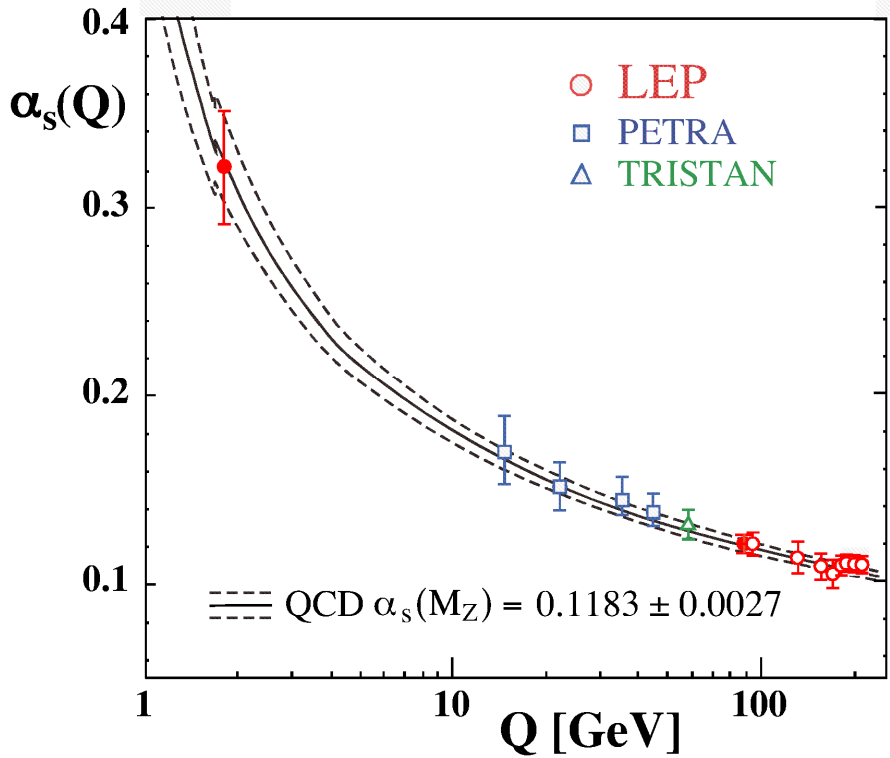}
\end{center}
\caption{\label{asq-LEP}
Summary of measurements of $\asq$ from LEP.
Results from $\epem$ annihilations at PETRA \cite{jadas,kluthas}
and TRISTAN (see \cite{concise}) are also included.
Open symbols are from event shapes in resummed NLO,
filled symbols from $\tau$ and $\z0$ hadronic decay
widths, in full NNLO QCD.
The curves represent the QCD predictions
of the running running coupling for the current world average 
of $\as$ \cite{as2002}.}
\end{figure}

The LEP measurements of $\as$, in the energy from $\mtau = 1.78$~GeV to 
$<\ecm > = 206$~GeV, are summarised in Figure~\ref{asq-LEP}, together with
earlier results from the TRISTAN collider (see \cite{concise}) 
and with recent results from
a \oq LEP-style"
re-analysis of PETRA data at lower c.m. energies \cite{jadas,kluthas}.
The data are compared to the QCD prediction of the running coupling constant,
calculated in $4^{th}$ order perturbation theory \cite{as-4loop} with
3-loop matching at the heavy quark pole masses \cite{3loop-matching}, for
the current world average value of $\amz = 0.1183 \pm 0.0027$ 
\cite{concise,as2002}\footnote{Note that this world average included
previous results of $\as$ from $\rz$ and from $\mtau$.}.
The specific energy dependence of $\as$ and the concept of Asymptotic Freedom
are strigently testified by the LEP results.

A combined value of $\amz$ from LEP data alone is calculated using the three
most significant results from $\tau$ decays, from $\rz$, 
both in complete NNLO QCD, and from the combined results from event 
shapes and jet production, using resummed NLO QCD predictions:
\begin{eqnarray}
\rm{\tau\ decays:} &\amz =& 0.1180 \pm 0.0030\ ,\nonumber \\
\rz: &\amz =& 0.1226 ^{+0.0058}_{-0.0038} \ ,
\ {\rm and}\nonumber \\
\rm{shapes:} &\amz =& 0.1202 \pm 0.0050 
\ .\nonumber
\end{eqnarray}
Since the errors are dominated by theoretical uncertainties 
which are largely correlated with each other, a combined value of $\amz$ is
calculated assuming an overall correlation factor between the three results
which is adjusted such that the total $\chi^2$ is unity per degree 
of freedom, giving
$$\amz = 0.1195 \pm 0.0034$$
for an overall correlation factor of 0.67,
as the final combined result from LEP.

\section{Colour Factors and nonabelian gauge structure of QCD}
\begin{figure}[htb]
\begin{center}
\epsfxsize12.0cm
\epsffile{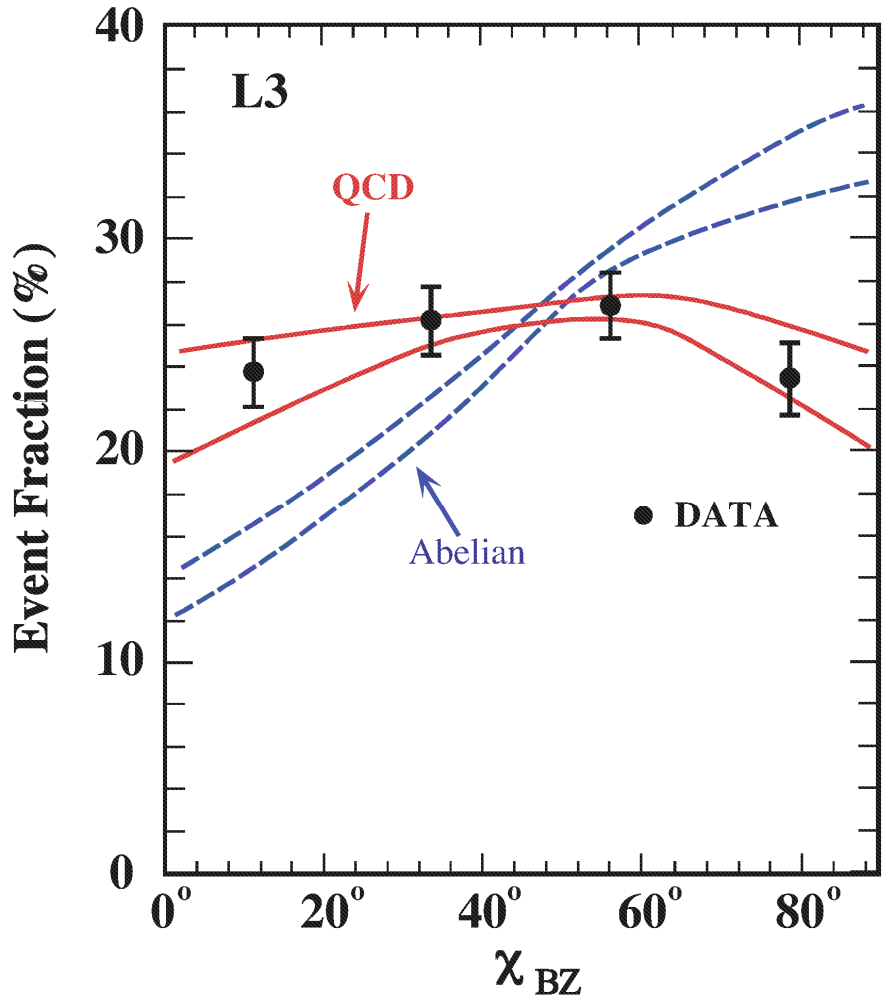}
\end{center}
\caption{\label{L3trip}
Distribution of the azimuthal angle between the planes spanned by
the two highest and the two lowest energetic jets
in 4-jet events measured at LEP \cite{l3-tgv}, 
together with predictions by QCD and by
an abelian \oq QED like" theory which does not include gluon self-coupling.}
\end{figure}

The central element giving rise to asymptotic freedom is
the gluon self-coupling in QCD which was studied in angular
correlations and energy distributions of 4-jet events.
The significance of such a measurement after one year of data 
taking at LEP is displayed in Figure~\ref{L3trip}~\cite{l3-tgv}.
Here, the distribution of the Bengtson-Zerwas angle \cite{bz} between the 
energy-ordered jet axes of reconstructed 4-jet events is 
compared with the predictions of QCD and with an Abelian theory where 
the gluon self-coupling does not exist.

\begin{figure}[htb]
\begin{center}
\epsfxsize12.0cm
\epsffile{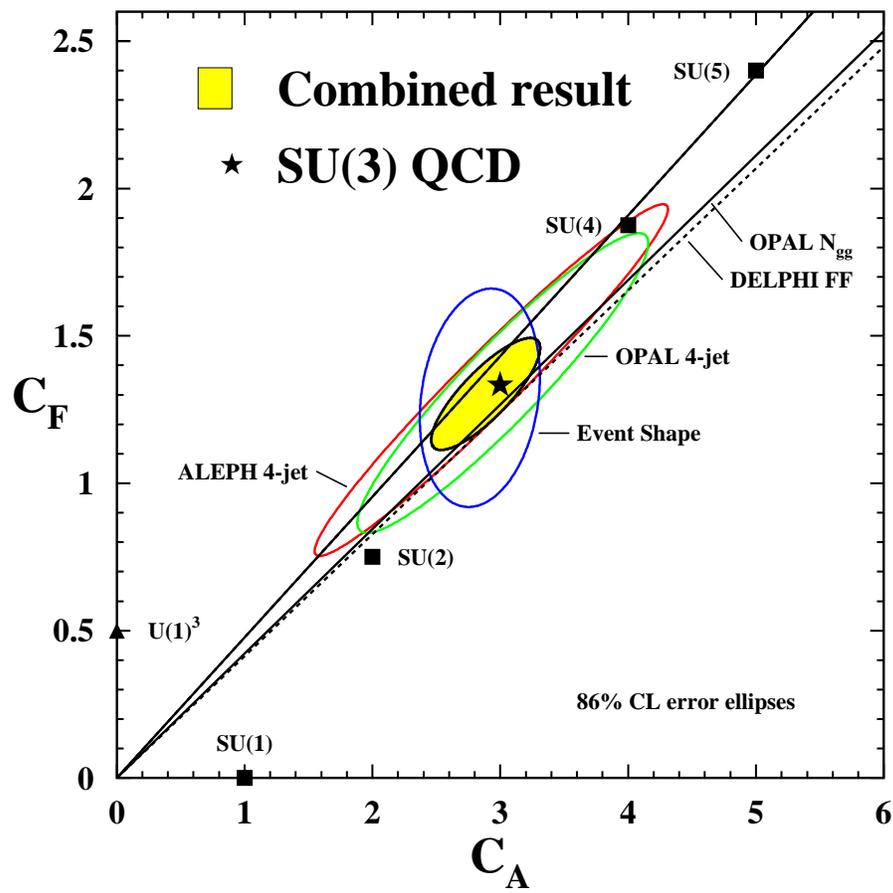}
\end{center}
\caption{\label{cacfplot}
Measurements and combination of the QCD colour factors $C_A$ and $C_F$
\cite{kluth-tgv}.}
\end{figure}

The current state-of-the art of such studies, which involve the 
analysis of several 4-jet angular correlations 
or fits to hadronic event shapes, is summarised \cite{kluth-tgv} 
in Figure~\ref{cacfplot}.
The data, with combined values of 
\begin{eqnarray}
C_A &=& 2.89 \pm 0.01\ {\rm (stat.)} \pm 0.21\ {\rm (syst.)} \\ \nonumber
C_F &=& 1.30 \pm 0.01\ {\rm (stat.)} \pm 0.09\ {\rm (syst.)}
\end{eqnarray}
are in excellent agreement with the gauge structure constants of 
QCD ($C_A \equiv N_C = 3$, $C_F = 4/3$ and $T_R = 1/2$), and rule out an
Abelian vector 
gluon model ($C_A = 0$, $C_F = 1$ and $T_R = 6$).
The existence of light colour-charged spin-1/2 supersymmetric 
partners of the gluon, the gluinos, is strongly disfavoured.

\section{Differences between q- and g-jets}
QCD predicts that quarks and gluons - due to their different colour charges 
- fragment differently: gluon initiated jets are 
expected to be broader than quark jets, the multiplicity of hadrons in 
gluon jets, $N_{had}$, should be larger than in quark jets, and particles in 
gluon jets are expected to be less energetic.

At LEP, corresponding studies at earlier $\epem$ colliders were further 
refined, e.g. by anti-tagging gluon jets through the help of high 
resolution silicon vertex detectors \cite{opal-g-tag}, 
by analysing gluon-inclusive jets recoiling against two 
other jets which are double-tagged to be a b-quark-antiquark system 
\cite{opal-gincl}, or by extracting the charged particle multiplicity of 
hypothetical gluon-gluon jet events from measurements of symmetric 3-jet 
events at LEP and from average hadronic 
(quark-antiquark-) events in $\epem$ annihilation \cite{delphi-qg}.

\begin{figure}[htb]
\begin{center}
\epsfxsize12.0cm
\epsffile{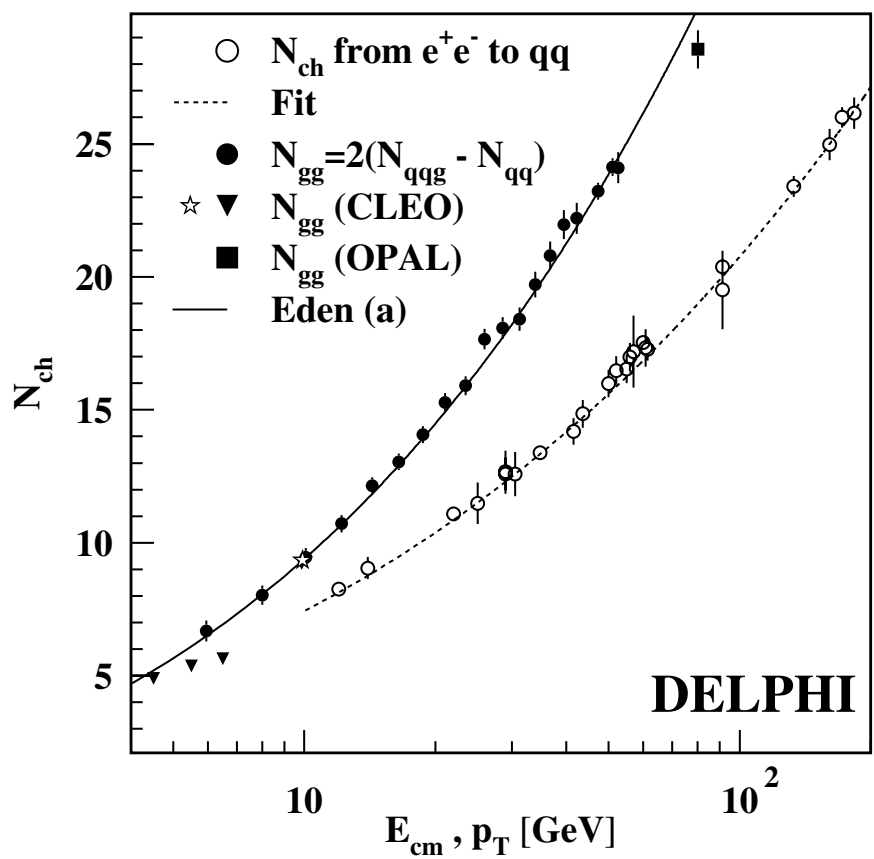}
\end{center}
\caption{\label{qg-mult}
Charged particle multiplicities for $gg$ and for $q\bar{q}$
final states as a function of the energy scale \cite{delphi-qg}.}
\end{figure}

One result of the latter type is displayed in Figure~\ref{qg-mult},
where the average charged particle multiplicities of gluon-gluon 
and of quark-antiquark configurations are compared to the
QCD predictions \cite{eden,webber-qq}.
These data, which confirm the QCD prediction of a higher colour charge 
of gluons compared to quarks, also provided a fit of the ratio 
$C_A/C_F = 2.22 \pm 0.11$ \cite{delphi-qg}, 
in perfect agreement with the QCD expectation of 2.25.

\section{QCD with Heavy Quarks}
\subsection{Gluon splitting into $c\bar{c}$ and $b\bar{b}$ quark pairs}
The fraction of $\epem \rightarrow {\rm hadrons}$ events in which a gluon splits
into a pair of heavy quarks, $c\bar{c}$ or $b\bar{b}$, is commonly referred to as
$\gcc$ and $\gbb$, respectively. 
These quantities are infrared safe, due to the cutoff by finite quark masses, 
and can therefore be calculated by means of perturbative QCD.
Such predictions, however, depend on the value of $\as$ as well as on the 
values of the quark masses. 
From leading and next-to-leading logarithmic approximations \cite{th-qsplit},
$\gcc$ is expected to be in the range of 1 percent 
and $\gbb$ to be about 1 permille.

Measurements of $\gcc$ and $\gbb$ are available by all LEP experiments 
as well as from the SLD experiment at the SLAC Linear Collider.
They are based on selections of 3-jet events with 
active tagging of two b-quarks, of two charmed mesons and/or of two leptons
in the gluon jet.
These measurements are summarised in Table~3.
Combining them results in
\begin{eqnarray}
\gcc &=& ( 3.05 \pm 0.14\ {\rm (exp.)} \pm 0.34\ {\rm (sys.)}) 10^{-2}\ {\rm and} \\
\gbb &=& ( 2.74 \pm 0.28\ {\rm (exp.)} \pm 0.72\ {\rm (sys.)}) 10^{-3}\ ,
\end{eqnarray}
where the experimental errors were combined in quadrature, the total
errors where determined by introducing a common correlation factor between all
measurements such that the overall $\chi^2$ per degree of freedom
adjusts to unity, and the systematic error is the quadratic difference of 
the latter two.
Without the result from SLD, the LEP results average to
$\gbb = ( 2.94 \pm 0.31 \pm 0.83) 10^{-3}$.

\subsection{Flavour independence of $\as$ and
  Measurements of the running b-quark mass}
Studies of the flavour dependence of $\as$ revealed a difference in jet 
rates and
event shapes between b quark and light quark events, of the order of a few 
percent (see ref. 4 in \cite{O-mb}).
These differences can be explained, in terms of NLO QCD calculations for massive
quarks \cite{bmass-th}, by effects of the large b-quark mass. 
With proper account of these effects, the flavour $in$dependence of $\as$, which
is a fundamental property of QCD, could be established within about 1~\% accuracy
for b-quarks, 4~\% for c-quarks and 5 to 10~\% for the light u-, d- and s-quarks,
see e.g. \cite{stenzel}.

\begin{figure}[htb]
\begin{center}
\epsfxsize12.0cm
\epsffile{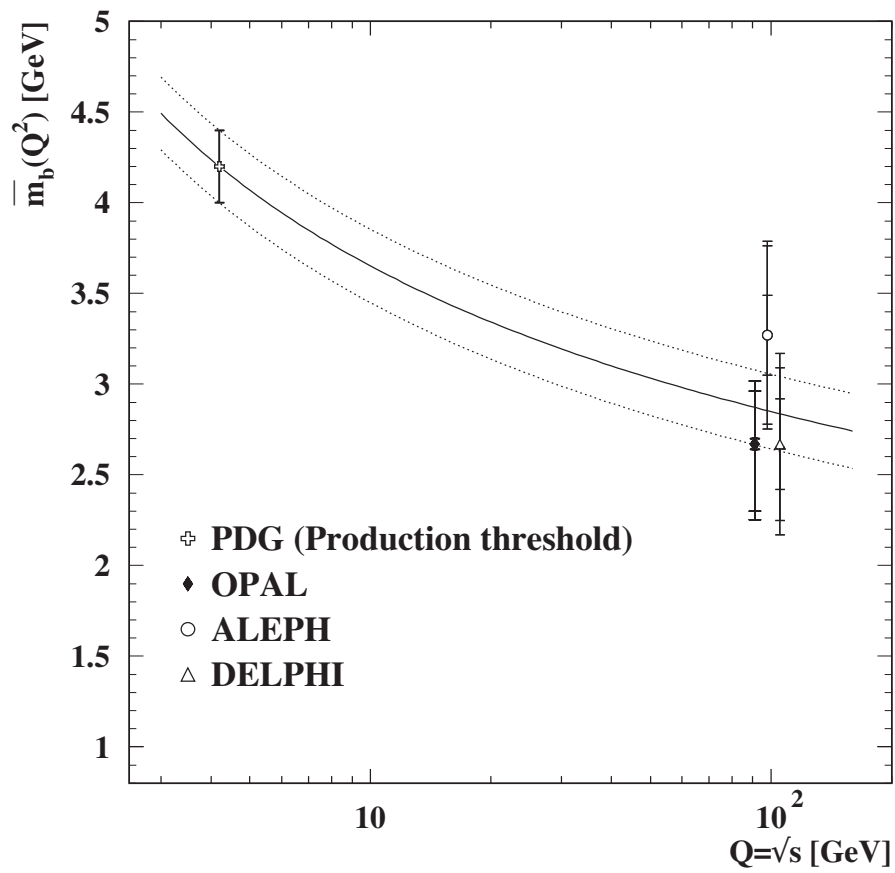}
\end{center}
\caption{\label{pr336_07}
Measurements of the b-quark mass at LEP, compared with the
value of $m_b$ at the bottom quark production threshold and the QCD expectation
of the running quark mass.}
\end{figure}

Taking the flavour independence of $\as$ for granted, the NLO QCD 
predictions for massive quarks can also be used to determine the b-quark 
mass at the energy scale of the $\z0$ boson.
QCD predicts that the quark masses depend on $\asq$ and thus are energy 
dependent, see e.g. \cite{mb-vermaseren}.
A summary of the measurements of the b-quark mass from LEP experiments
\cite{D-mb,A-mb,O-mb}
is given in Figure~\ref{pr336_07}.
Also shown is the QCD prediction for the running b-quark mass, normalised 
to its value at the production threshold, ${\rm m_b (m_b) = (4.2 \pm 0.2)\ GeV}$
\cite{pdg2000}, and using the world average value of $\amz = 0.1184 \pm 0.0031$ 
\cite{concise}.

Combining the LEP measurements with the same treatment of (correlated) errors as
described in the previous subsection, results in
$${\rm m_b (\mz) = (2.82 \pm 0.02\ (stat.) \pm 0.37\ (sys.))\ GeV}\ ,$$ 
which excludes a constant b-quark mass with a significance of 3.3 standard
deviations.

\section{Two photon physics}
\begin{figure}[htb]
\begin{center}
\epsfxsize12.0cm
\epsffile{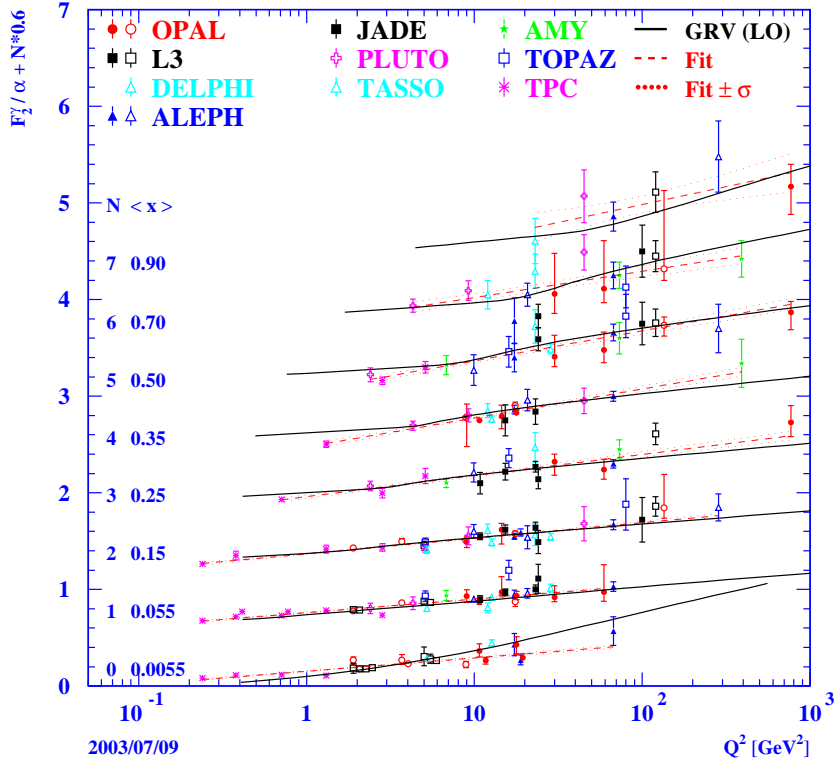}
\end{center}
\caption{\label{F2-x-Q}
Compilation of measurements of the hadronic photon structure function
$F_2^{\gamma}$ in $\epem$ collisions \cite{nisius-pr}.}
\end{figure}

Extensive studies of two-photon scattering processes leading to hadronic 
final states have been performed at LEP; 
for summary reports on this particular topic see e.g. 
\cite{nisius-pr,rembold,nisius-02}.
Scaling violations are seen in a compilation of measurements of
the photon hadronic structure function
$F_2^{\gamma} (x,Q^2)$ from LEP and from previous $\epem$ experiments
\cite{nisius-pr}, see Figure~\ref{F2-x-Q}.
The LEP data, especially those obtained at LEP-II, extend the range
of measurements of $F_2^{\gamma}$ to $<Q^2>$ up to 780~GeV$^2$,
the largest scale of photon structure probed in $\epem$
collisions.

\begin{figure}[htb]
\begin{center}
\epsfxsize12.0cm
\epsffile{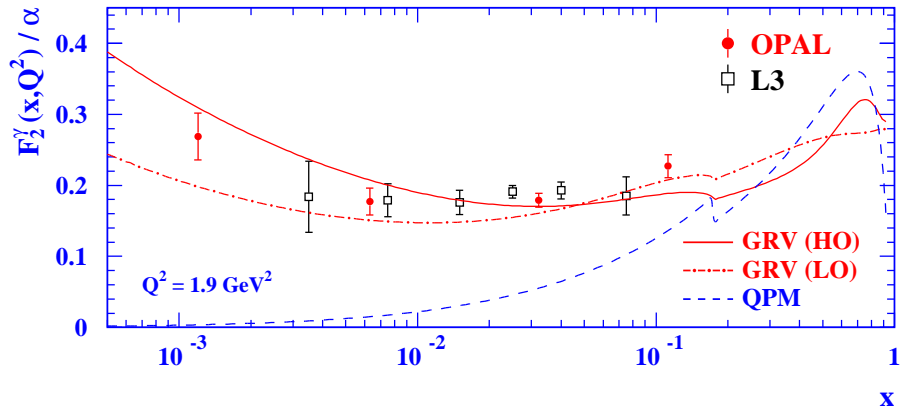}
\end{center}
\caption{\label{f2lowx}
Measurements of $F_2^{\gamma}$ at small $Q^2$ and small $x$ \cite{nisius-pr}.}
\end{figure}

LEP measurements also extend the range of data at very small $x$, down to
$x \sim 10^{-3}$, as seen in Figure~\ref{f2lowx}.  
The data are compatible with a rise of  $F_2^{\gamma}$ 
as predicted by leading (LO) and
higher order (HO) perturbative QCD \cite{GRV}, while the
simple quark-parton model (QPM) is naturally inadequate to describe 
data in this regime.

\section{Summary and conclusions}
The successful running of LEP has 
led to a significant increase of knowledge about hadron production and the 
dynamics of quarks and gluons at high energies.
Precise determinations of $\as$ at the smallest and the largest c.m.
energies available to date, superior treatment and evaluation of
experimental and theoretical uncertainties,
experimental confirmation of asymptotic freedom 
and of the gluon self coupling, detailed studies of differences between 
quark and gluon jets, verification of the running b-quark mass and of the 
flavour independence of $\as$, deeper understanding of power corrections and 
of hadronisation models to describe the nonperturbative hadronisation 
domain, and detailed studies of hadronic systems in two-photon scattering 
processes were summarised in this report, proving QCD as a consistent 
theory which accurately describes the phenomenology of the Strong 
Interaction.

Future developments in this field are within reach: NNLO QCD 
calculations and predictions for jet and event shape observables
will soon be available; they will initiate further analyses of the LEP
data which will provide 
even more accurate and more detailed determinations of $\as$.

\vspace{5 mm}
\noindent {\bf Acknowledgements.}
The scientific results summarised in this review are achieved by the coherent
work of a huge number of scientists, engineers and technicians, at CERN as
well at all the institutes who participated in the LEP program worldwide.
Special thanks go to CERN and the groups running the LEP collider so
efficiently, and also to the large number 
of funding agencies who had the breath to support the project through all 
these years of planning, of constructing and of running LEP.
I am indebted to S. Kluth, to R. Nisius, to P. Zerwas 
and to the LEPQCDWG for their 
inspiring inputs, and for allowing to use their material in this review.

%

\renewcommand{\arraystretch}{1.3}
\begin{table}[htb]
\caption{
Typical numbers of hadronic events obtained by each of the four
LEP experiments, at and around the principal c.m. energies.
Numbers for $\ecm \ge 161$~GeV are corrected for and do not include
4-fermion final states. }
\begin{center}
  {
\begin{tabular}{|c|r|r|r|r|r|r|r|r|}
   \hline
$\ecm$ [GeV] & 91.2 & 133 & 161 & 172 & 183 & 189 & 200 & 206 \\
\# of events &
$4\times 10^6$ & 800 &  300 &  200 & 1200 & 3000 & 3000 & 3000 \\
\hline
\end{tabular} }
\end{center} 
\end{table}

\renewcommand{\arraystretch}{1.3}
\begin{table}[htb]
\caption{
Combined results of $\as (Q)$ for major LEP c.m. 
energies $Q$ \cite{lepqcdwg2}. }
\begin{center}
{ 
\begin{tabular}{|c|c c c c c|c|}
   \hline
$Q$  & $\as (Q)$ & stat. & exp. & hadr. & theory & total \\
(GeV) & & error &  error &  error & error & error  \\
\hline
91.2  & 0.1199 & $\pm 0.0002$ & $\pm 0.0008$ & $\pm 0.0017$ &
        $^{+0.0048}_{-0.0047}$ & $^{+0.0052}_{-0.0051}$ \\
133.0 & 0.1135 & $\pm 0.0016$ & $\pm 0.0012$ & $\pm 0.0013$ &
        $^{+0.0045}_{-0.0044}$ & $^{+0.0051}_{-0.0050}$ \\
161.0 & 0.1081 & $\pm 0.0025$ & $\pm 0.0015$ & $\pm 0.0011$ & 
        $\pm 0.0041$          &  $\pm 0.0051$ \\
172.0 & 0.1049 & $\pm 0.0029$ & $\pm 0.0017$ & $\pm 0.0009$ & 
        $\pm 0.0040$ &  $\pm 0.0053$ \\
183.0 & 0.1077 & $\pm 0.0013$ & $\pm 0.0009$ & $\pm 0.0008$ & 
        $^{+0.0037 }_{-0.0038 }$ &  $^{+0.0041 }_{-0.0042 }$ \\
189.0 & 0.1092 & $\pm 0.0008$ & $\pm 0.0009$ & $\pm 0.0008$ & 
        $^{+0.0037 }_{-0.0038 }$ &  $^{+0.0040 }_{-0.0041 }$ \\
200.0 & 0.1080 & $\pm 0.0009$ & $\pm 0.0010$ & $\pm 0.0007$ & 
        $^{+0.0036 }_{-0.0037 }$ &  $^{+0.0039 }_{-0.0040 }$ \\
206.0 & 0.1078 & $\pm 0.0009$ & $\pm 0.0008$ & $\pm 0.0007$ & 
        $^{+0.0033 }_{-0.0035 }$ &  $^{+0.0036 }_{-0.0038 }$ \\
\hline
\end{tabular} }
\end{center} 
\end{table}

\renewcommand{\arraystretch}{1.3}
\begin{table}[htb]
\caption{
Compilation of results on fractions of gluons splitting 
into $c\bar{c}$ and $b\bar{b}$ . }
\begin{center}
  {
\begin{tabular}{|l|c|c|c|}
   \hline
exp.  & $\gcc \times 10^{-2}$ & $\gbb \times 10^{-3}$ & ref. \\
\hline
ALEPH  & $3.26 \pm 0.23 \pm 0.42$ & $2.77 \pm 0.42 \pm 0.57$ & \cite{A-gcc,A-gbb} \\
DELPHI & --- & $3.3 \pm 1.0 \pm 0.8$ & \cite{D-gbb} \\
L3     & $2.45 \pm 0.29 \pm 0.53$ & --- & \cite{L-gcc} \\
OPAL   & $3.20 \pm 0.21 \pm 0.38$ & $3.07 \pm 0.53 \pm 0.97$ & \cite{O-gcc,O-gbb} \\
SLD    & --- & $2.44 \pm 0.59 \pm 0.34$ & \cite{S-gbb} \\
\hline
\end{tabular} }
\end{center} 
\end{table}

\end{document}